# A New Variant of Benes Network : Its Topological Characterisation and Comparative Analysis

Parvez Ali[1], Annmaria Baby[2,*], D. Antony Xavier[2], Eddith Sarah Varghese[2], Theertha Nair A.[2], Haidar Ali[3]

[2]Department of Mechanical Engineering, College of Engineering, Qassim University, Buraydah, 51452, Saudi Arabia.

[2]Department of Mathematics, Loyola College, University of Madras, Chennai, India.

[3]Department of Mathematics, Riphah International University, Faisalabad, Pakistan.

Mail: p.ali@qu.edu.sa, 21dmt001@loyolacollege.edu*, antonyxavier@loyolacollege.edu, 19dmt001@loyolacollege.edu, 21dmt005@loyolacollege.edu, haidar3830@gmail.com

**Abstract:** The modern era always looks into advancements in technology. Design and topology of interconnection networks play a mutual role in development of technology. Analysing the topological properties and characteristics of an interconnection network is not an easy task. Graph theory helps in solving this task analytically and efficiently through the use of numerical parameters known as distance based topological descriptors. These descriptors have considerable applications in various fields of computer science, chemistry, biology, etc. This paper deals with the evaluation of topological descriptors for an n-dimensional multistage interconnection network, the benes network, $BB(n)$. Also, a new variant of interconnection network is derived from the benes network, named as augmented benes network and denoted as $BB^*(n)$. The topological descriptors for the benes derived network are also determined in this work. Further, the benes network and augmented benes network undergoes a comparative analysis based on few network parameters, which helps to understand the efficiency of newly derived benes network. A broadcasting algorithm for the augmented benes network is also provided.

# 1 Introduction

Networking was first made use in 1950s, in telecommunications to connect phone calls through switchboards that switched between electric connections or switches to make the connections. In the computing world, networking provides a method for fast communication between multiple processors of a computer and between multiple computers connected to a network. The network is composed of links and switches. A single processor, a cluster of processors, or a set of memory modules can make up each node in the network, which is made up of interconnections and switches that help transfer data from the source to the destination. A network is defined by its layout, transit scheme, switching mechanism, and diffusion strategy. Topology is indeed the arrangement used to link individual switches to other components such as processors, memories and other switches. Data interchange among processors in a parallel system is also made possible via network. The concept behind interconnection networks is that when a computer activity that requires a wide range of data cannot be adequately handled by a single processor, the task is divided into parallel tasks that are executed simultaneously, greatly reducing the time consumption. Higher transmission speed between various components of parallel system requires effective interconnectivity networks. The layout of the switch and the wiring pattern in between the switches determine the overall configuration of the network. The degree of the switch, routing technique, and its internal buffer determines the topology and routing algorithm of the networks. The several network kinds include direct connection networks, indirect connection networks, bus networks, multistage connections and crossbar switches.

The design of an interconnection network and analysing its efficiency is an important research problem in parallel computing. In the field of computer science, average distance is a principal parameter, that is used to measure the communication cost of networks. Researchers have also developed various other parameters to analyse and differentiate interconnection network models. These parameters include degree, diameter, message traffic density, network throughput, graph density, total connectivity, packing density and network cost. A network is said to be efficient if it has low and constant degree, minimum diameter, minimum average distance, low message traffic density, high network throughput, high graph density, high total connectivity, low network cost and high packing density [1].

The interconnection networks can be tailored into a graph with vertices representing the processors and edges representing the communication links. Even, graphs can be used to design new interconnection networks. From the graph models of the networks, efficiently one can characterise its topological properties. Therefore, it is purposeful to study the topological descriptors of networks using the graph models. The idea of topological descriptors was first introduced by H. Wiener in 1947 [2]. Eventually many other topological descriptors were introduced following the pioneering works of Wiener [2] and Randić [3]. The parameter average distance is nothing, but the Wiener index itself. These descriptors are related with distinct properties of a structure and hence, can be used to investigate the relationships between the structure, its properties and activities. Topological descriptors have marked its practical presence in chemistry, informatics, biology, etc. [2-7].

The butterfly network is one of the notable multistage interconnection network, which is mainly used to perform fast Fourier transform [8] and other appealing applications are given in [9]. The $n$-dimensional butterfly is denoted as $BF(n)$ and it consists of $2^n$ $(n+1)$ nodes arranged in $n+1$ levels, such that each level has $2^n$ nodes. Each node of $BF(n)$ can be assigned with a label $(j,k)$, where $j$ is a $n$-bit binary number that denotes the column of the node and $k$ is the level of the node ranging from 0 to $n$. Two nodes $(j,k)$ and $(j',k')$ of $BF(n)$ are linked, if $k' = k+1$ and either $j$ and $j'$ are identical or $j$ and $j'$ differ precisely in $k^{th}$ bit. The benes network is back to back butterflies, whose $n^{th}$ dimensional network is denoted as $BB(n)$ and it consists of $2^n(2n+1)$ nodes arranged in $2n+1$ levels. In $BB(n)$, each node can be labelled as $(j,k)$ where $j$ is a $n$-bit binary number and $0 \leq k \leq 2n$. Two nodes of $BB(n)$, $(j,k)$ and $(j',k')$ are linked, if $k' = k+1$ and either $j$ and $j'$ are identical or $j$ and $j'$ differ precisely in $k^{th}$ bit for $0 \leq k \leq n-1$ and they differ in $(2n-k-1)^{th}$ bit for $n \leq k \leq 2n$. Benes network is a multi-stage interconnection network, which has considerable applications and is known for permutation routing [8,10,11]. In topological aspect, each node is considered as a vertex and the links between them as edges. Manual et al. [8] has described the diamond representation for butterfly network as well as benes network together with their vertex labelling. The 3-dimensional benes network, $BB(3)$ is presented in Figure 1.

In this paper, a new variant of benes network, named as augmented benes network is derived. The $n$-dimensional augmented benes network is denoted as $BB^*(n), n \geq 1$ consists of $2^n(2n+1)$ nodes arranged in $2n+1$ levels and two nodes $(j,k)$ and $(j',k')$ are linked, if the following conditions are satisfied.
  i. $k' = k+1$ and $j,j'$ are identical or $j,j'$ differ in $k^{th}$ bit for $0 \leq k \leq n-1$ and they differ in $(2n-k-1)^{th}$ bit for $n \leq k \leq 2n$.
  ii. $k' = k$ and $j,j'$ differ in $k^{th}$ bit or $j,j'$ differ precisely in $(k-1)^{th}$ bit for $0 \leq k \leq n$ and for $n+1 \leq k \leq 2n$ they differ in $(2n-k)^{th}$ or $(2n-k-1)^{th}$ bit.

Hence, the number of vertices of the n-dimensional augmented benes network remains the same as that of the $n$-dimensional benes network, whereas the number of edges increases to $2^{n-1}(12n-1)$. The 3-dimensional augmented benes network, $BB^*(3)$ is illustrated in Figure 2, which provides a better constructional clarity.

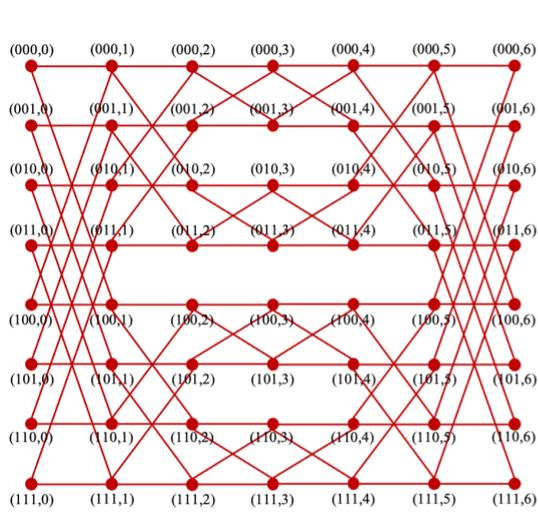
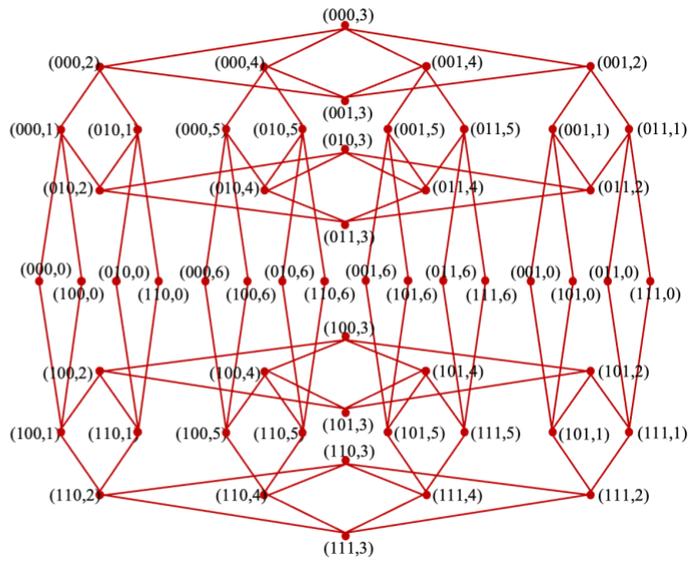

Figure 1: The 3-dimensional benes network, $BB(3)$.

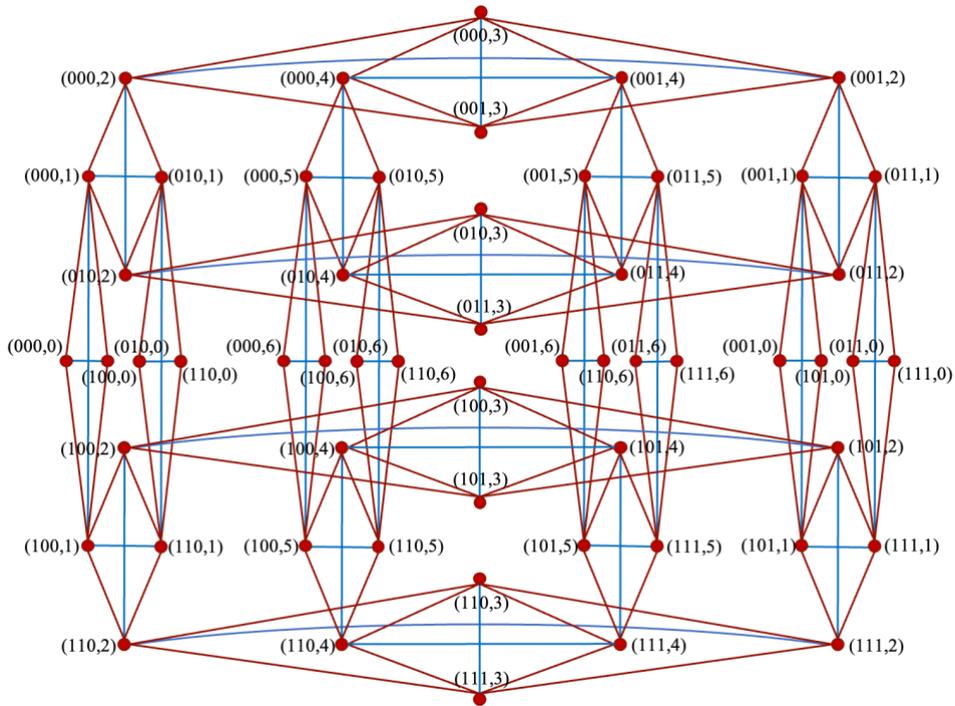

Figure 2: The 3-dimensional augmented benes network, $BB^*(3)$.

Profound researchers have studied the distance based and degree based topological descriptors for butterfly networks [12-15]. Literature shows study on degree based descriptors for benes network [12-14,16] but not on distance based descriptors. In this paper, this research gap is vanished by determining the distance based topological descriptors for benes network, $BB(n)$. Additionally, a new network is derived from the benes network, denoted as $BB^*(n)$ and the topological descriptors for the same are studied. Also the benes network and augmented benes network are analysed with the help of topological network parameters and a comparison study is carried out inorder to understand the efficiency of newly derived benes network.

## 2 Mathematical Terminologies

Consider a simple connected graph $\zeta$ with vertex set, $V(\zeta)$ and edge set, $E(\zeta)$. The number of edges incident to a vertex $\mu$ is the degree of that vertex and is denoted as $d\mu$. $D_\zeta(\mu, \eta)$ denotes the shortest-path distance between the vertices $\mu$ and $\eta$ in $\zeta$. The neighborhood of a vertex $\mu$ is denoted as $N_\zeta(\mu)$, which is the set of vertices adjacent to $\mu$ or it is a subgraph of $\zeta$ induced by all the vertices adjacent to $\mu$. If the vertex $\mu$ is included in the set or induced subgraph, it is said to be closed neighborhood and is denoted as $N_\zeta[\mu]$. For more basic definitions refer to [17].

The concepts of isometric subgraph, partial cubes, convex subgraph and Djoković-Winkler ($\Theta$) relation are to be recollected while working on topological indices [18]. The Djoković-Winkler relation is a major concept applied in computing topological indices. For two edges $\varepsilon = \mu\eta$ and $\rho = \kappa\nu$ of $\zeta$, if $d_\zeta(\eta, \nu) + d_\zeta(\mu, \kappa) \neq d_\zeta(\mu, \nu) + d_\zeta(\eta, \kappa)$, then we say $\varepsilon$ is related to $\rho$. The relation $\Theta$ is reflexive, symmetric and transitive in case of partial cubes. Its transitive closure $\Theta^*$ forms an equivalence relation in general and partitions the edge set into convex components. Let $F = \{F_1, F_2, \ldots, F_x\}$ be the $\Theta^*$ equivalence class. A partition $\mathfrak{E} = \{\xi_1, \xi_2, \ldots, \xi_\delta\}$ of $E(G)$ is said to be coarser than partition $F$ if each set $\xi_i$ is the union of one or more $\Theta^*$-classes of $G$. The concept of strength weighted graph is presented in [18]. A strength weighted graph is denoted as $\zeta_{sw} = (\zeta, [w_v, s_v], s_e)$, whose vertex weight, vertex strength and edge strength are denoted as $w_v, s_v$ and $s_e$ respectively. For convenience, let us consider strength-weighted graph $\zeta_{sw}$ as $\zeta$.

For an edge $\varepsilon = \mu\eta$, the sets
$N_\mu(\varepsilon|\zeta) = \{\lambda \in \mathbb{V}(\zeta) : d_\zeta(\mu, \lambda) < d_\zeta(\eta, \lambda)\}$ and $M_\mu(\varepsilon|\zeta) = \{\chi \in \mathbb{E}(\zeta) : d_\zeta(\mu, \chi) < d_\zeta(\eta, \chi)\}$.

The cardinality of sets $N_\mu(\varepsilon|\zeta)$ and $M_\mu(\varepsilon|\zeta)$ are described as

$n_\mu(\varepsilon|\zeta) = \sum_{\lambda \in N_\mu(\varepsilon|\zeta)} w_v(\lambda)$ and $m_\mu(\varepsilon|\zeta) = \sum_{\lambda \in N_\mu(\varepsilon|\zeta)} s_v(\lambda) + \sum_{\chi \in M_\mu(\varepsilon|\zeta)} s_e(\chi)$

respectively and the values of $n_\eta(\varepsilon|\zeta)$ and $m_\eta(\varepsilon|\zeta)$ are analogous. For more terminologies and definitions refer to [18]. The distance based molecular descriptors [18] and their corresponding mathematical expression for strength-weighted graph $\zeta$ are given in Table 1.

**Table 1:** Distance based structural descriptors for strength-weighted graph $\zeta$.

| Molecular Descriptor | Mathematical Expression |
| --- | --- |
| Wiener | $W(\zeta) = \sum_{\{\mu,\eta\} \subseteq \mathbb{V}(\zeta)} w_v(\mu) w_v(\eta) d_\zeta(\mu, \eta)$ |
| Szeged | $Sz_v(\zeta) = \sum_{\varepsilon=\mu\eta \in \mathbb{E}(\zeta)} s_e(\varepsilon) n_\mu(\varepsilon|\zeta) n_\eta(\varepsilon|\zeta)$ |
| Edge-Szeged | $Sz_e(\zeta) = \sum_{\varepsilon=\mu\eta \in \mathbb{E}(\zeta)} s_e(\varepsilon) m_\mu(\varepsilon|\zeta) m_\eta(\varepsilon|\zeta)$ |
| Edge-Vertex-Szeged | $Sz_{ev}(\zeta) = \frac{1}{2} \sum_{\varepsilon=\mu\eta \in \mathbb{E}(\zeta)} s_e(\varepsilon)[n_\mu(\varepsilon|\zeta) m_\eta(\varepsilon|\zeta) + m_\mu(\varepsilon|\zeta) n_\eta(\varepsilon|\zeta)]$ |
| Padmakar-Ivan | $PI(\zeta) = \sum_{\varepsilon=\mu\eta \in \mathbb{E}(\zeta)} s_e(\varepsilon) [m_\mu(\varepsilon|\zeta) + m_\eta(\varepsilon|\zeta)]$ |
| Mostar | $Mo(G) = \sum_{\varepsilon=\mu\eta \in \mathbb{E}(\zeta)} s_e(\varepsilon) |n_\mu(\varepsilon|\zeta) - n_\eta(\varepsilon|\zeta)|$ |
| Edge-Mostar | $Mo_e(G) = \sum_{\varepsilon=\mu\eta \in \mathbb{E}(\zeta)} s_e(\varepsilon) |m_\mu(\varepsilon|\zeta) - m_\eta(\varepsilon|\zeta)|$ |

**Theorem 1:** *Let $\zeta$ be a simple connected weighted graph. Let $X_i, 1 \leq i \leq a$ and $Y_j, 1 \leq j \leq b$ be subgraphs of $\zeta$ such that $X_i, Y_j \cong K_m, m \geq 2 \; \forall \; i,j$, with $w_v(\mu) = \alpha \; \forall \; \mu \in \mathbb{V}(X_i)$ and $w_v(\eta) = \beta, \forall \; \eta \in \mathbb{V}(Y_j)$. If*

i. *Closed neighborhood of a vertex $\mu \in \mathbb{V}(X_i)$, $N_\zeta[\mu]$ is a subgraph induced by the vertex set $\cup_{j=1}^b \mathbb{V}(Y_j) \cup \mathbb{V}(X_i)$ and $N_\zeta[\mu] \cong N_\zeta[\eta]$, for $\mu \in \mathbb{V}(X_i), \eta \in \mathbb{V}(X_h), 1 \leq i, h \leq a$.*
ii. *Closed neighborhood of a vertex $\mathfrak{p} \in \mathbb{V}(Y_j)$, $N_\zeta[\mathfrak{p}]$ is a subgraph induced by the vertex set $\cup_{j=1}^a \mathbb{V}(X_i) \cup \mathbb{V}(Y_j)$ and $N_\zeta[\mathfrak{p}] \cong N_\zeta[\mathfrak{q}]$, for $\mathfrak{p} \in \mathbb{V}(Y_j), \mathfrak{q} \in \mathbb{V}(Y_h), 1 \leq i, h \leq b$.*

*Then,*
$$W(\zeta) = m^2(a(a-1)\alpha^2 + b(b-1)\beta^2 + ab\alpha\beta) + \binom{m}{2}(a\alpha^2 + b\beta^2).$$

**Proof:** The graph $\zeta$ under the given conditions can be depicted as given in Figure 3. As an example, the graph $\zeta$ with $m = 3, a = 3$ and $b = 2$ is presented in Figure 4.

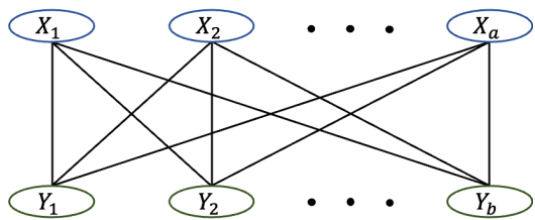

**Figure 3:** The graph $\zeta$, where $\zeta$, where $X_i, Y_j \cong K_m; 1 \leq i \leq a, 1 \leq j \leq b, m \geq 2$.

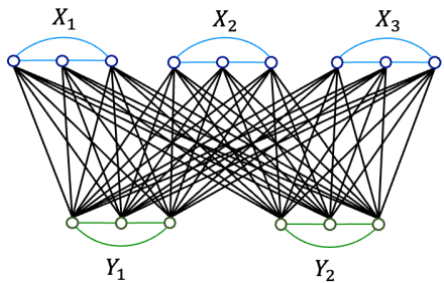

**Figure 4:** The graph $\zeta$, where $\zeta$, where $X_i, Y_j \cong K_3; 1 \leq i \leq 3, 1 \leq j \leq 2$.

By definition, $W(\zeta) = \sum_{\{\mu,\eta\} \subseteq \mathbb{V}(\zeta)} w_v(\mu) w_v(\eta) d_\zeta(\mu, \eta)$. To determine the wiener index of $\zeta$, the vertex set $\{\mu, \eta\} \subseteq \mathbb{V}(\zeta)$ can be chosen from three cases.

- *Case 1:* $\mu \in \mathbb{V}(X_i), \eta \in \mathbb{V}(X_j); 1 \leq i,j \leq a$.
- *Case 2:* $\mu \in \mathbb{V}(Y_i), \eta \in \mathbb{V}(Y_j); 1 \leq i,j \leq b$.
- *Case 3:* $\mu \in \mathbb{V}(X_i), \eta \in \mathbb{V}(Y_j); 1 \leq i \leq a, 1 \leq j \leq b$.

It can be noted that, under Case 1 and Case 2, distance between the vertices $\mu$ and $\eta$ will be one, when $i = j$ and it will be two, when $i \neq j$. Also, under Case 3, distance between the vertices $\mu$ and $\eta$ will be one for any value of $i$ and $j$.

For complete graph $K_m$, we have the result, $W(K_m) = \binom{m}{2}$. Hence the wiener index for the weighted complete graphs $X_i, 1 \leq i \leq a$ and $Y_j, 1 \leq j \leq b$, which are isomorphic to $K_m$, can be given as, $W(X_i) = \binom{m}{2} \alpha^2$ and $W(Y_j) = \binom{m}{2} \beta^2$.

By using the definition for wiener index given in Table 1 and the observed data, the wiener index for the given graph $\zeta$,

$$W(\zeta) = \sum_{\{\mu,\eta\}\subseteq \mathbb{V}(\zeta)} w_v(\mu)\, w_v(\eta)\, d_\zeta(\mu,\eta)$$

$$= \frac{1}{2} \sum_{\mu\in\mathbb{V}(X_p); \eta\in\mathbb{V}(X_q)} w_v(\mu) w_v(\eta) d_\zeta(\mu,\eta)$$

$$+ \frac{1}{2} \sum_{\mu\in\mathbb{V}(Y_r); \eta\in\mathbb{V}(Y_s)} w_v(\mu) w_v(\eta) d_\zeta(\mu,\eta) + \frac{1}{2} \sum_{\mu\in\mathbb{V}(X_i); \eta\in\mathbb{V}(Y_j)} w_v(\mu) w_v(\eta) d_\zeta(\mu,\eta);$$

where, $1 \le p,q,i \le a;\; 1 \le r,s,j \le b$

$$= \frac{1}{2}\left( \sum_{\mu\in\mathbb{V}(X_p);\eta\in\mathbb{V}(X_p)} w_v(\mu)w_v(\eta)d_\zeta(\mu,\eta) + \sum_{\mu\in\mathbb{V}(X_p);\eta\in\mathbb{V}(X_p), p\ne q} w_v(\mu)w_v(\eta)d_\zeta(\mu,\eta) \right)$$

$$+ \frac{1}{2}\left( \sum_{\mu\in\mathbb{V}(Y_r);\eta\in\mathbb{V}(Y_r)} w_v(\mu)w_v(\eta)d_\zeta(\mu,\eta) + \sum_{\mu\in\mathbb{V}(Y_r);\eta\in\mathbb{V}(Y_s), r\ne s} w_v(\mu)w_v(\eta)d_\zeta(\mu,\eta) \right)$$

$$+ \frac{1}{2}\left( \sum_{\mu\in\mathbb{V}(X_i);\eta\in\mathbb{V}(Y_j)} w_v(\mu)w_v(\eta)d_\zeta(\mu,\eta) \right)$$

$$= a \binom{m}{2}\alpha^2 + a(a-1)m^2\alpha^2 + b\binom{m}{2}\beta^2 + b(b-1)m^2\beta^2 + abm^2\alpha\beta$$

$$= m^2(a(a-1)\alpha^2 + b(b-1)\beta^2 + ab\alpha\beta) + \binom{m}{2}(a\alpha^2 + b\beta^2).$$

## 3 Distance Based Structural Descriptors

The distance based structural descriptors for the benes network $BB(n), n \ge 2$, and augmented benes network, $BB^*(n), n \ge 2$ are computed using the quotient graph approach. This method is used, since regular cut method is not applicable for the augmented benes network, since they are not partial cubes. Here, the technique is to convert the original graph into quotient graphs using $\Theta^*$-classes and further, the descriptors of each quotient graphs are determined and added up correspondingly in order to obtain the descriptors of original graph. This transformation reduces the original graphs into compact graphs, making the computation easier and faster.

### 3.1 Benes Network, $BB(n)$

The benes network, $BB(n), n \ge 2$, has $(2n+1)2^n$ vertices and $n2^{n+2}$ edges. Using the Djoković-Winkler relation, the $\Theta^*$classes of $BB(n), n \ge 2$ were determined and based on it, $BB(n)$ is reduced to $n$ quotient graphs, $G/\xi_\delta, 1 \le \delta \le n$. The quotient graphs are isomorphic to complete bipartite graphs, $K_{(2^{n-\delta+1}, 2^{\delta+1})}; n \ge 2, 1 \le \delta \le n$. For example, the quotient graphs for $BB(2), G/\xi_1, G/\xi_2$ are isomorphic to $K_{(4,4)}$ and $K_{(2,8)}$ respectively. The $\Theta^*$-classes $\xi_1$ and $\xi_2$ chosen for $BB(2)$ are illustrated in Figure 5. The generalised quotient graph, $G/\xi_\delta, 1 \le \delta \le n$, for $BB(n), n \ge 2$ is presented in Figure 6.

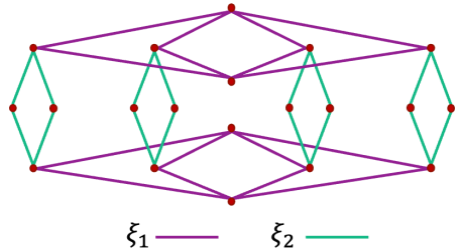

**Figure 5:** The $\Theta^*$-classes $\xi_1$ and $\xi_2$ chosen for $BB(2)$.

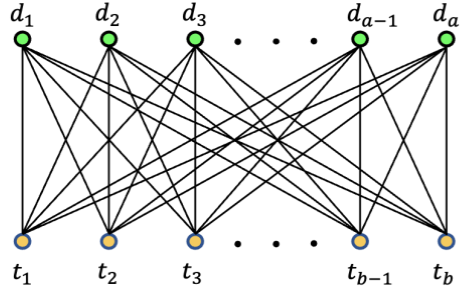

- $(w_v(d_a), s_v(d_a)) = (2^{n-\delta}(n-\delta+1), 2^{n-\delta+1}(n-\delta)), 1 \le a \le 2^{\delta+1}$
- $(w_v(t_b), s_v(t_b)) = (2^{\delta-1}(2\delta-1), 2^{\delta+1}(\delta-1)), 1 \le b \le 2^{n-\delta+1}$

**Figure 6:** The quotient graph of $BB(n)$, $G/\xi_\delta, 1 \le \delta \le n$.

**Theorem 2:** *For $n \ge 2$,*
$$W(BB(n)) = \frac{2^n}{3}(50n \times 2^n - 69 \times 2^n - 9n^2 \times 2^n + 10n^3 \times 2^n + 69)$$

**Proof:** The Wiener index for $BB(n), n \ge 2$, is computed with the help of reduction theorem [15]. In this process, the quotient graphs of the benes network are converted to reduced graphs and further reduction theorem is applied to determine the result. The reduced graph of the generalised quotient graph, $G/\xi_\delta, 1 \le \delta \le n$, for $BB(n), n \ge 2$ is presented in Figure 7.

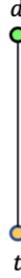

$w_v(d) = 2^{n+1}(n-\delta+1), w_v(t) = 2^n(2\delta-1)$

**Figure 7:** The reduced graph of the generalised quotient graph for $BB(n), n \ge 2$.

Using the reduction theorem ([15] - Theorem 2.1),
$$W(BB(n)/\xi_\delta) = 2^{n+1}(n-\delta+1)(2^n)(2\delta-1) + 2\binom{2^{\delta+1}}{2}\left(2^{n-\delta}(n-\delta+1)\right)^2$$
$$+ 2\binom{2^{n-\delta+1}}{2}\left(2^{\delta-1}(2\delta-1)\right)^2$$

$$W(BB(n)) = \sum_{\delta=1}^{n} W(BB(n)/\xi_\delta)$$

$$= \frac{2^n}{3}(50n \times 2^n - 69 \times 2^n - 9n^2 \times 2^n + 10n^3 \times 2^n + 69)$$

**Theorem 3:** *For $n \ge 2$,*
$$Sz(BB(n)) = \frac{2 \times 2^n}{27}\{(2608 - 1308n + 288n^2 + 36n^3) \times 2^{2n}$$
$$- (2700 + 423n - 27n^2 - 18n^3) \times 2^n + 92\}$$

$$Sz_e\big(BB(n)\big) = 2^n \left\{ \left( \frac{20096}{27} - \frac{2864}{9}n + \frac{112}{3}n^2 + \frac{32}{3}n^3 \right) \times 2^{2n} \right.$$
$$\left. - \left( 792 + \frac{496}{3}n + 16n^2 - \frac{16}{3}n^3 \right) \times 2^n + 4n + \frac{1288}{27} \right\}$$

$$Sz_{ev}\big(BB(n)\big) = 2^n \left\{ \left( \frac{10636}{27} - \frac{1660}{9}n + \frac{92}{3}n^2 + \frac{16}{3}n^3 \right) \times 2^{2n} \right.$$
$$\left. - \left( 408 + \frac{242}{3}n + 2n^2 - \frac{8}{3}n^3 \right) \times 2^n + \frac{380}{27} \right\}$$

$$PI\big(BB(n)\big) = 2^n \{(16n^2 - 16n + 24) \times 2^n - 8n - 24\}$$

$$Mo\big(BB(n)\big) = \sum_{\delta=1}^{n} 2^{n+2} \left| \left( \left(2^{n-\delta}(n - \delta + 1)\right) + \left(2^{n-\delta+1} - 1\right)\left(2^{\delta-1}(2\delta - 1)\right) \right) \right.$$
$$\left. - \left( \left(2^{\delta+1} - 1\right)\left(2^{n-\delta}(n - \delta + 1)\right) + \left(2^{\delta-1}(2\delta - 1)\right) \right) \right|$$

$$Mo_e\big(BB(n)\big) = \sum_{\delta=1}^{n} 2^{n+2} \left| \left( \left(2^{n-\delta+1}(n - \delta)\right) + \left(2^{n-\delta+1} - 1\right)\left(2^{\delta+1}(\delta - 1)\right) + 2^{n-\delta+1} - 1 \right) \right.$$
$$\left. - \left( \left(2^{\delta+1} - 1\right)\left(2^{n-\delta+1}(n - \delta)\right) + \left(2^{\delta+1}(\delta - 1)\right) + 2^{\delta+1} - 1 \right) \right|$$

**Proof:** The quotient graph, $BB(n)/\xi_\delta$, $1 \leq \delta \leq n$ for $BB(n)$, $n \geq 2$ is given in Figure 6. Since $BB(n)/\xi_\delta$, $1 \leq \delta \leq n$ is isomorphic to complete bipartite graph, $K_{(2^{n-\delta+1}, 2^{\delta+1})}$, the cardinality of its edge set is $2^{n+2}$.

For each edge $\epsilon = \mu\eta \in BB(n)/\xi_\delta$, the neighborhood sets $N_\mu(\varepsilon|BB(n)/\xi_\delta)$, $N_\eta(\varepsilon|BB(n)/\xi_\delta)$, $M_\mu(\varepsilon|BB(n)/\xi_\delta)$ and $M_\eta(\varepsilon|BB(n)/\xi_\delta)$ and their corresponding cardinalities are determined. It is observed that, for each edge, these values remain same and are obtained as follows.

$$n_\mu(\varepsilon|(BB(n)/\xi_\delta)) = \left( \left(2^{n-\delta}(n - \delta + 1)\right) + \left(2^{n-\delta+1} - 1\right)\left(2^{\delta-1}(2\delta - 1)\right) \right).$$
$$n_\eta(\varepsilon|(BB(n)/\xi_\delta)) = \left( \left(2^{\delta+1} - 1\right)\left(2^{n-\delta}(n - \delta + 1)\right) \right) + \left(2^{\delta-1}(2\delta - 1)\right).$$
$$m_\mu(\varepsilon|(BB(n)/\xi_\delta)) = \left( \left(2^{n-\delta+1}(n - \delta)\right) + \left(2^{n-\delta+1} - 1\right)\left(2^{\delta+1}(\delta - 1)\right) + 2^{n-\delta+1} - 1 \right).$$
$$m_\eta(\varepsilon|(BB(n)/\xi_\delta)) = \left( \left(2^{\delta+1} - 1\right)\left(2^{n-\delta+1}(n - \delta)\right) + \left(2^{\delta+1}(\delta - 1)\right) + 2^{\delta+1} - 1 \right).$$

The Szeged indices, Padmakar-Ivan index and Mostar indices of the network are computed by applying these results in the expressions from Table 1 as follows.

$$Sz_v(BB(n)/\xi_\delta) = \sum_{\varepsilon=\mu\eta\in\mathbb{E}(BB(n)/\xi_\delta)} s_e(\varepsilon) \, n_\mu(\varepsilon|(BB(n)/\xi_\delta)) n_\eta(\varepsilon|(BB(n)/\xi_\delta))$$
$$= 2^{n+2} \left\{ \left( \left(2^{n-\delta}(n - \delta + 1)\right) + \left(2^{n-\delta+1} - 1\right)\left(2^{\delta-1}(2\delta - 1)\right) \right) \right.$$
$$\left. \times \left( \left(2^{\delta+1} - 1\right)\left(2^{n-\delta}(n - \delta + 1)\right) + \left(2^{\delta-1}(2\delta - 1)\right) \right) \right\}$$

$$\sum_{\delta=1}^{n} Sz_v(BB(n)/\xi_\delta)$$
$$= \frac{2 \times 2^n}{27}\{(2608 - 1308n + 288n^2 + 36n^3) \times 2^{2n}$$
$$- (2700 + 423n - 27n^2 - 18n^3) \times 2^n + 92\}$$

$$Sz_e(BB(n)/\xi_\delta) = \sum_{\varepsilon=\mu\eta\in\mathbb{E}(BB(n)/\xi_\delta)} s_e(\varepsilon)\, m_\mu(\varepsilon|\,(BB(n)/\xi_\delta))m_\eta(\varepsilon|\,(BB(n)/\xi_\delta))$$

$$= 2^{n+2}\{\left(\left(2^{n-\delta+1}(n-\delta)\right) + \left(2^{n-\delta+1} - 1\right)\left(2^{\delta+1}(\delta-1)\right) + 2^{n-\delta+1} - 1\right)$$
$$\times \left(\left(2^{\delta+1} - 1\right)\left(2^{n-\delta+1}(n-\delta)\right) + \left(2^{\delta+1}(\delta-1)\right) + 2^{\delta+1} - 1\right)\}$$

$$\sum_{\delta=1}^{n} Sz_e(BB(n)/\xi_\delta)$$
$$= 2^n\left\{\left(\frac{20096}{27} - \frac{2864}{9}n + \frac{112}{3}n^2 + \frac{32}{3}n^3\right) \times 2^{2n}\right.$$
$$\left. - \left(792 + \frac{496}{3}n + 16n^2 - \frac{16}{3}n^3\right) \times 2^n + 4n + \frac{1288}{27}\right\}$$

$$Sz_{ev}(BB(n)/\xi_\delta)$$
$$= \frac{1}{2}\sum_{\varepsilon=\mu\eta\in\mathbb{E}(BB(n)/\xi_\delta)} s_e(\varepsilon)\left[n_\mu(\varepsilon|\,(BB(n)/\xi_\delta))m_\eta(\varepsilon|\,(BB(n)/\xi_\delta))\right.$$
$$\left. + m_\mu(\varepsilon|\,(BB(n)/\xi_\delta))n_\eta(\varepsilon|\,(BB(n)/\xi_\delta))\right]$$
$$= 2^{n+2}\left[\{\left(\left(2^{n-\delta}(n-\delta+1)\right) + \left(2^{n-\delta+1} - 1\right)\left(2^{\delta-1}(2\delta-1)\right)\right)\right.$$
$$\times \left(\left(2^{\delta+1} - 1\right)\left(2^{n-\delta+1}(n-\delta)\right) + \left(2^{\delta+1}(\delta-1)\right) + 2^{\delta+1} - 1\right)\}$$
$$+ \{\left(\left(2^{\delta+1} - 1\right)\left(2^{n-\delta}(n-\delta+1)\right) + \left(2^{\delta-1}(2\delta-1)\right)\right)$$
$$\left.\times \left(\left(2^{n-\delta+1}(n-\delta)\right) + \left(2^{n-\delta+1} - 1\right)\left(2^{\delta+1}(\delta-1)\right) + 2^{n-\delta+1} - 1\right)\}\right]$$

$$\sum_{\delta=1}^{n} Sz_{ev}(BB(n)/\xi_\delta)$$
$$= 2^n\left\{\left(\frac{10636}{27} - \frac{1660}{9}n + \frac{92}{3}n^2 + \frac{16}{3}n^3\right) \times 2^{2n}\right.$$
$$\left. - \left(408 + \frac{242}{3}n + 2n^2 - \frac{8}{3}n^3\right) \times 2^n + \frac{380}{27}\right\}$$

$$PI(BB(n)/\xi_\delta) = \sum_{\varepsilon=\mu\eta\in\mathbb{E}(BB(n)/\xi_\delta)} s_e(\varepsilon)\, m_\mu(\varepsilon|\,(BB(n)/\xi_\delta)) + m_\eta(\varepsilon|\,(BB(n)/\xi_\delta))$$
$$= 2^{n+2}\{\left(\left(2^{n-\delta+1}(n-\delta)\right) + \left(2^{n-\delta+1} - 1\right)\left(2^{\delta+1}(\delta-1)\right) + 2^{n-\delta+1} - 1\right)$$
$$+ \left(\left(2^{\delta+1} - 1\right)\left(2^{n-\delta+1}(n-\delta)\right) + \left(2^{\delta+1}(\delta-1)\right) + 2^{\delta+1} - 1\right)\}$$

$$\sum_{\delta=1}^{n} PI(BB(n)/\xi_\delta) = 2^n\{(16n^2 - 16n + 24) \times 2^n - 8n - 24\}$$

$$Mo(BB(n)/\xi_\delta) = \sum_{\varepsilon=\mu\eta \in \mathbb{E}(BB(n)/\xi_\delta)} s_e(\varepsilon) \, |n_\mu(\varepsilon|\,(BB(n)/\xi_\delta)) - n_\eta(\varepsilon|\,(BB(n)/\xi_\delta))|$$

$$= 2^{n+2} \left| \left( \left(2^{n-\delta}(n-\delta+1)\right) + \left(2^{n-\delta+1} - 1\right)\left(2^{\delta-1}(2\delta-1)\right) \right) \right.$$
$$\left. - \left( \left(2^{\delta+1} - 1\right)\left(2^{n-\delta}(n-\delta+1)\right) + \left(2^{\delta-1}(2\delta-1)\right) \right) \right|$$

$$\sum_{\delta=1}^{n} Mo(BB(n)/\xi_\delta)$$

$$= \sum_{\delta=1}^{n} 2^{n+2} \left| \left( \left(2^{n-\delta}(n-\delta+1)\right) + \left(2^{n-\delta+1} - 1\right)\left(2^{\delta-1}(2\delta-1)\right) \right) \right.$$
$$\left. - \left( \left(2^{\delta+1} - 1\right)\left(2^{n-\delta}(n-\delta+1)\right) + \left(2^{\delta-1}(2\delta-1)\right) \right) \right|$$

$$Mo_e(BB(n)/\xi_\delta) = \sum_{\varepsilon=\mu\eta \in \mathbb{E}(BB(n)/\xi_\delta)} s_e(\varepsilon) \, |m_\mu(\varepsilon|\,(BB(n)/\xi_\delta)) - m_\eta(\varepsilon|\,(BB(n)/\xi_\delta))|$$

$$= 2^{n+2} \left| \left( \left(2^{n-\delta+1}(n-\delta)\right) + \left(2^{n-\delta+1} - 1\right)\left(2^{\delta+1}(\delta-1)\right) + 2^{n-\delta+1} - 1 \right) \right.$$
$$\left. - \left( \left(2^{\delta+1} - 1\right)\left(2^{n-\delta+1}(n-\delta)\right) + \left(2^{\delta+1}(\delta-1)\right) + 2^{\delta+1} - 1 \right) \right|$$

$$\sum_{\delta=1}^{n} Mo_e(BB(n)/\xi_\delta)$$

$$= \sum_{\delta=1}^{n} 2^{n+2} \left| \left( \left(2^{n-\delta+1}(n-\delta)\right) + \left(2^{n-\delta+1} - 1\right)\left(2^{\delta+1}(\delta-1)\right) + 2^{n-\delta+1} - 1 \right) \right.$$
$$\left. - \left( \left(2^{\delta+1} - 1\right)\left(2^{n-\delta+1}(n-\delta)\right) + \left(2^{\delta+1}(\delta-1)\right) + 2^{\delta+1} - 1 \right) \right|$$

Hence,

$$Sz_v\big(BB(n)\big) = \sum_{\delta=1}^{n} Sz_v(BB(n)/\xi_\delta)$$

$$= \frac{2 \times 2^n}{27}\{(2608 - 1308n + 288n^2 + 36n^3) \times 2^{2n} - (2700 + 423n - 27n^2 - 18n^3) \times 2^n + 92\}$$

$$Sz_e\big(BB(n)\big) = \sum_{\delta=1}^{n} Sz_e(BB(n)/\xi_\delta)$$

$$= 2^n \left\{ \left(\frac{20096}{27} - \frac{2864}{9}n + \frac{112}{3}n^2 + \frac{32}{3}n^3\right) \times 2^{2n} - \left(792 + \frac{496}{3}n + 16n^2 - \frac{16}{3}n^3\right) \times 2^n + 4n + \frac{1288}{27} \right\}$$

$$Sz_{ev}(BB(n)) = \sum_{\delta=1}^{n} Sz_{ev}(BB(n)/\xi_\delta)$$

$$= 2^n \left\{ \left( \frac{10636}{27} - \frac{1660}{9}n + \frac{92}{3}n^2 + \frac{16}{3}n^3 \right) \times 2^{2n} - \left( 408 + \frac{242}{3}n + 2n^2 - \frac{8}{3}n^3 \right) \times 2^n + \frac{380}{27} \right\}$$

$$PI(BB(n)) = \sum_{\delta=1}^{n} PI(BB(n)/\xi_\delta) = 2^n\{(16n^2 - 16n + 24) \times 2^n - 8n - 24\}$$

$$Mo(BB(n)) = \sum_{\delta=1}^{n} Mo(BB(n)/\xi_\delta)$$
$$= \sum_{\delta=1}^{n} 2^{n+2} \left| \left( \left(2^{n-\delta}(n-\delta+1)\right) + \left(2^{n-\delta+1} - 1\right)\left(2^{\delta-1}(2\delta - 1)\right) \right) \right.$$
$$\left. - \left( \left(2^{\delta+1} - 1\right)\left(2^{n-\delta}(n-\delta+1)\right) + \left(2^{\delta-1}(2\delta - 1)\right) \right) \right|$$

$$Mo_e(BB(n)) = \sum_{\delta=1}^{n} Mo_e(BB(n)/\xi_\delta)$$
$$= \sum_{\delta=1}^{n} 2^{n+2} \left| \left( \left(2^{n-\delta+1}(n-\delta)\right) + \left(2^{n-\delta+1} - 1\right)\left(2^{\delta+1}(\delta - 1)\right) + 2^{n-\delta+1} - 1 \right) \right.$$
$$\left. - \left( \left(2^{\delta+1} - 1\right)\left(2^{n-\delta+1}(n-\delta)\right) + \left(2^{\delta+1}(\delta - 1)\right) + 2^{\delta+1} - 1 \right) \right|$$

### 3.2 Augmented Benes Network, $BB^*(n)$

The augmented benes network, $BB^*(n), n \geq 2$, has $(2n+1)2^n$ vertices and $(12n-1)2^{n-1}$ edges. Using the Djoković-Winkler relation, the $\Theta^*$-classes of $BB^*(n), n \geq 2$, were determined and based on those classes, $BB^*(n)$ can be reduced to n quotient graphs, $BB^*(n)/\xi_\sigma, 1 \leq \sigma \leq n$. For example, $BB^*(2)$ can have two quotient graphs, $BB^*(2)/\xi_1$, and $BB^*(2)/\xi_2$. The $\Theta^*$-classes $\xi_1$ and $\xi_2$ chosen for $BB^*(2)$ and their quotient graphs are depicted in Figure 8, Figure 9 and Figure 10 respectively. The quotient graphs for $BB^*(n), n \geq 2$, can be standardised into two types, $BB^*(n)/\xi_1$, and $BB^*(n)/\xi_\sigma, 2 \leq \sigma \leq n$ as illustrated in Figure 11 and Figure 12 respectively.

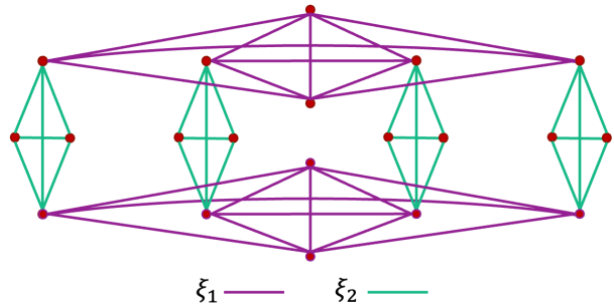

$\xi_1$ ——  $\xi_2$ ——

**Figure 8:** The $\Theta^*$ −classes $\xi_1$ and $\xi_2$ chosen for $BB^*(2)$.

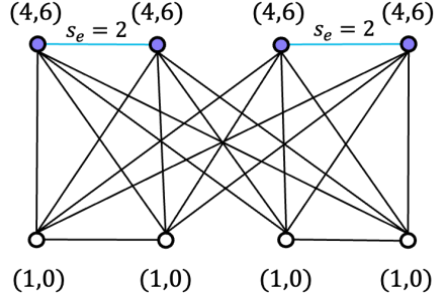

**Figure 9:** The quotient graph of $BB^*(2)$, $BB^*(2)/\xi_1$.

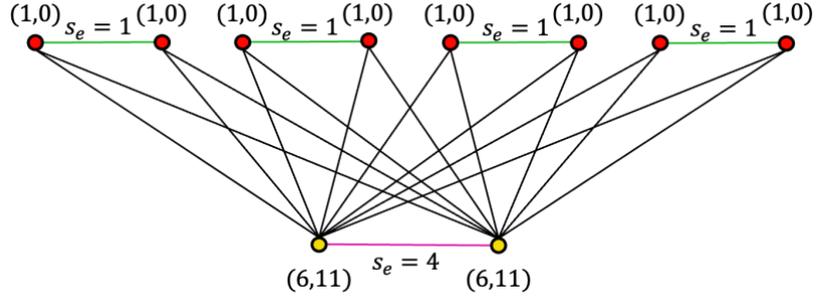

**Figure 10:** The quotient graph of $BB^*(2)$, $BB^*(2)/\xi_2$.

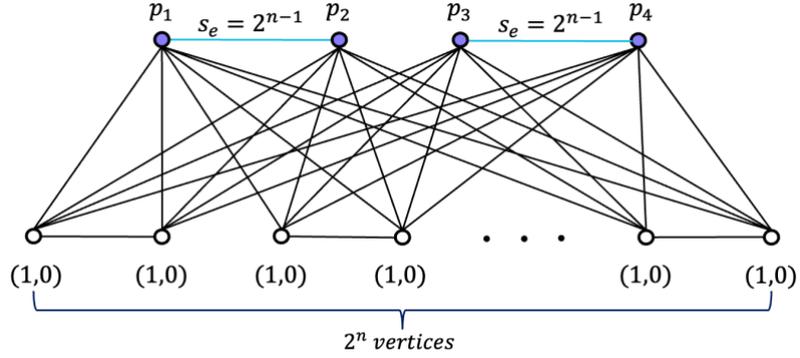

○ $(w_v(p), s_v(p)) = (w_v(p_c), s_v(p_c)) = (n\,2^{n-1}, 3(n-1)\,2^{n-1}), 1 \leq c \leq 4$

**Figure 11:** The first type of quotient graph for $BB^*(n)$, $BB^*(n)/\xi_1$.

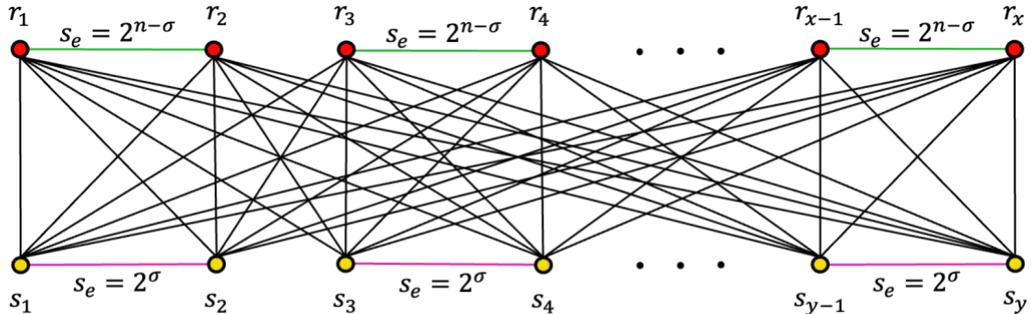

● $(w_v(r), s_v(r)) = (w_v(r_x), s_v(r_x)) = ((n-\sigma+1)\,2^{n-\sigma}, 3(n-\sigma)\,2^{n-\sigma}), 1 \leq x \leq 2^{\sigma+1}$

○ $(w_v(s), s_v(s)) = (w_v(s_y), s_v(s_y)) = ((2\sigma-1)\,2^{\sigma-1}, (12\sigma-13)\,2^{\sigma-2}), 1 \leq y \leq 2^{n-\sigma+1}$

**Figure 12:** The second type of quotient graph for $BB^*(n)$, $BB^*(n)/\xi_\sigma, 2 \leq \sigma \leq n$.

**Theorem 4.** For $n \geq 2$,
$$W(BB^*(n)) = \frac{2^n}{6}\left((20n^3 - 36n^2 + 148n - 207) \times 2^n + 207\right)$$

**Proof:** It can be observed that, the quotient graphs $BB^*(n)/\xi_1$, and $BB^*(n)/\xi_\sigma$ are isomorphic to $\zeta$ defined in Theorem 1 under specific parameters.

• Consider quotient graph $BB^*(n)/\xi_1$ (Figure 11).

The graph $BB^*(n)/\xi_1$ is isomorphic to $\zeta$ with $a = 2, b = 2^{n-1}$ and $m = 2$ defined in Theorem 1. That is, $X_1, X_2$ and $Y = Y_1, Y_2, \ldots, Y_{2^{n-1}}$ are subgraphs of $BB^*(n)/\xi_1$ such that $X_i, 1 \leq i \leq 2$ and $Y_j, 1 \leq j \leq 2^{n-1}$ are isomorphic to complete graph $K_2$. Also $\alpha = n2^{n-1}$ and $\beta = 1$. Hence the wiener index of graph $BB^*(n)/\xi_1$ can be determined by applying Theorem 1.

$$W(BB^*(n)/\xi_1) = 2^2\left(2(2-1)(n\,2^{n-1})^2 + 2^{n-1}(2^{n-1}-1)(1)^2 + 2 \times 2^{n-1}(n\,2^{n-1})(1)\right)$$
$$+ \binom{2}{2}\left(2(n\,2^{n-1})^2 + 2^{n-1}(1)^2\right)$$

$$= 2^n\left\{\left(\frac{5}{2}n^2 + 2n + 1\right)2^n - \frac{3}{2}\right\}$$

• Consider quotient graph $BB^*(n)/\xi_\sigma, 2 \leq \sigma \leq n$ (Figure 12).

The graph $BB^*(n)/\xi_\sigma$ is isomorphic to $\zeta$ defined in Theorem 1 with $a = 2^{n-\sigma-1}, b = 2^{\sigma-1}$ and $m = 2$. Therefore, $X_1, X_2, \ldots, X_{2^{n-\sigma-1}}, Y_1, Y_2, \ldots, Y_{2^{\sigma-1}}$ are subgraphs of $BB^*(n)/\xi_\sigma$, such that $X_i \cong K_2$, $1 \leq i \leq 2^{n-\sigma-1}$ and $Y_j \cong K_2, 1 \leq j \leq 2^{\sigma-1}$. Also $\alpha = (n - \sigma + 1)2^{n-\sigma}$ and $\beta = (2\sigma - 1)2^{\sigma-1}$. Hence the wiener index of graph $BB^*(n)/\xi_\sigma$ can be determined by applying Theorem 1.

$$W(BB^*(n)/\xi_\sigma) = 2^2\left(2^{n-\sigma-1}(2^{n-\sigma-1}-1)\left((n-\sigma+1)2^{n-\sigma}\right)^2\right.$$
$$+ 2^{\sigma-1}(2^{\sigma-1}-1)\left((2\sigma-1)2^{\sigma-1}\right)^2$$
$$\left.+ 2^{n-\sigma-1} \times 2^{\sigma-1} \times (n-\sigma+1)2^{n-\sigma} \times (2\sigma-1)2^{\sigma-1}\right)$$
$$+ \binom{m}{2}\left(2^{n-\sigma-1}\left((n-\sigma-1)2^{n-\sigma}\right)^2 + 2^{\sigma-1}\left((2\sigma-1)2^{\sigma-1}\right)^2\right)$$
$$= 2^{2n-\sigma}(2^{\sigma+2}-3)(n-\sigma-1)^2 + 2^{2n+1}(2\sigma-1)(n-\sigma+1) - 2^{n-2}(2\sigma-1)^2(3 \times 2^\sigma - 2^{n+2})$$

$$\sum_{\sigma=1}^{n} W(BB^*(n)/\xi_\sigma) = \frac{2^n}{6}\left((20n^3 - 51n^2 + 136n - 213)2^n + 216\right)$$

Hence, $\quad W(BB^*(n)) = W(BB^*(n)/\xi_1) + \sum_{\sigma=2}^{n} W(BB^*(n)/\xi_\sigma)$

$$= \frac{2^n}{6}\left((20n^3 - 36n^2 + 148n - 207) \times 2^n + 207\right)$$

**Theorem 5:** For $n \geq 2$,
$$Sz(BB^*(n)) = \frac{2^n}{54}\{(15988 - 7212n + 882n^2 + 144n^3) \times 2^{2n}$$
$$- (16632 + 3366n - 270n^2 - 180n^3) \times 2^n + 617\}$$

$$Sz_e(BB^*(n)) = 2^n\left\{\left(\frac{30709}{12} - 922n + 57n^2 + 24n^3\right) \times 2^{2n}\right.$$
$$\left.- \left(2756 + \frac{1445}{2}n + \frac{123}{2}n^2 - 30n^3\right) \times 2^n + 4n + \frac{587}{3}\right\}$$

$$Sz_{ev}(BB^*(n)) = 2^n \left\{ \left( \frac{16145}{18} - \frac{1087}{3}n + 34n^2 + 8n^3 \right) \times 2^{2n} \right.$$
$$\left. - \left( 943 + \frac{909}{4}n + \frac{11}{4}n^2 - 10n^3 \right) \times 2^n + \frac{410}{9} \right\}$$

$$PI(BB^*(n)) = 2^n \{(73 - 36n + 24n^2) \times 2^n - 8n - 78\}$$

$$Mo(BB^*(n)) = 2^{n+2} |n2^{n-1} - 2^n + 3|$$
$$+ \sum_{\sigma=2}^{n} 2^{n+2} |(2^{n-\sigma}(n - \sigma + 1) + (2\sigma - 1)(2^n - 2^\sigma) - 2^{\sigma-1}(2\sigma - 1)$$
$$- 2^{n-\sigma}(2^{\sigma+1} - 2)(n - \sigma + 1))|$$

$$Mo_e(BB^*(n)) = 2^{n+2} |2^{n-1} - 3(n-1)2^{n-1} + 2^n - 6|$$
$$+ \sum_{\sigma=2}^{n} 2^{n+2} |((36n - 36\sigma + 16) 2^{n-\sigma-2} - (24n - 48\sigma + 26) 2^{n-2}$$
$$- (36\sigma - 23) 2^{\sigma-2})|$$

**Proof:** There are two types of quotient graphs for $BB^*(n), n \geq 2$, namely $BB^*(n)/\xi_1$ and $BB^*(n)/\xi_\sigma, 2 \leq \sigma \leq n$ as depicted in Figure 11 and Figure 12 respectively.

The first type of quotient graph, $BB^*(n)/\xi_1$ has $9 \times 2^{n-1} + 2$ edges and the second type of quotient graph, $BB^*(n)/\xi_\sigma, 2 \leq \sigma \leq n$ has $2^{2n+2}$ edges. As in Theorem 3, the neighborhood sets and their corresponding cardinality for each edge in both quotient graphs are determined and applied in the mathematical expressions from Table 1 to compute the Szeged, Padmakar-Ivan and Mostar indices. As a result, the following outcomes are obtained.

• For quotient graph $BB^*(n)/\xi_1$ (Figure 11), $w_v(p) = n2^{n-1}, s_v(p) = 3(n-1)2^{n-1}$

$$Sz(BB^*(n)/\xi_1) = 2^{n+2}(w_v(p) + 2^n - 2)(2w_v(p) + 1) + 2^n(w_v(p))^2 + 2^{n-1}$$

$$= \frac{2^n}{4} \{(16n + 9n^2)2^{2n} + (16 - 24n)2^n - 30\}$$

$$Sz_e(BB^*(n)/\xi_1) = 2^{n+2} \{(s_v(p) + 2^{n+1} - 2)(2s_v(p) + 4 + 2^{n-1})\}$$
$$+ 2^n \{(s_v(p) + 2^n)(s_v(p) + 2^n)\} + 2^{n-1}\{4 \times 4\}$$
$$= \frac{2^n}{4} \{(81n^2 - 42n - 19)2^{2n} + 112 \times 2^n - 96\}$$

$$Sz_{ev}(BB^*(n)/\xi_1)$$
$$= \frac{1}{2} \left\{ 2^{n+2} \{((w_v(p) + 2^n - 2)(2s_v(p) + 4 + 2^{n-1})) \right.$$
$$+ ((2w_v(p) + 1)(s_v(p) + 2^{n+1} - 2))\} + 2^n \{2(w_v(p)(s_v(p) + 2^n))\}$$
$$\left. + 2^{n-1}\{4 + 4\} \right\}$$

$$= \frac{2^n}{4} \{(27n^2 + 17n - 20)2^{2n} + (76 - 36n)2^n - 72\}$$

$$PI(BB^*(n)/\xi_1) = 2^{n+2}\{(s_v(p) + 2^{n+1} - 2) + (2s_v(p) + 4 + 2^{n-1})\}$$
$$+ 2^n\{(s_v(p) + 2^n) + (s_v(p) + 2^n)\} + 2^{n-1}\{4 + 4\}$$
$$= 3 \times 2^n\big((7n - 3)2^n + 4\big)$$

$$Mo(BB^*(n)/\xi_1) = 2^{n+2}|(w_v(p) + 2^n - 2) - (2w_v(p) + 1)| + 2 \times 0 + 2^{n-1} \times 0$$
$$= 2^{n+2}|n2^{n-1} - 2^n + 3|$$

$$Mo(BB^*(n)/\xi_1) = 2^{n+2}|(s_v(p) + 2^{n+1} - 2) - (2s_v(p) + 4 + 2^{n-1})| + 2 \times 0 + 2^{n-1} \times 0$$
$$= 2^{n+2}|2^{n-1} - 3(n-1)2^{n-1} + 2^n - 6|$$

- For quotient graph $BB^*(n)/\xi_\sigma, 2 \leq \sigma \leq n$ (figure 12), the similar steps used in computing the results for quotient graph $BB^*(n)/\xi_1$, are implemented, and the following results are obtained.

$$Sz_v(BB^*(n)/\xi_\sigma)$$
$$= 2^{n+2}\{(2^{n-\sigma}(n - \sigma + 1) - (2^\sigma - 2^n)(2\sigma - 1))(2^{\sigma-1}(2\sigma - 1)$$
$$+ 2^{n-\sigma}(n - \sigma + 1)(2^{\sigma+1} - 2))\}$$
$$+ 2^{n-\sigma} \times 2^\sigma\{(2^{n-\sigma}(n - \sigma + 1))(2^{n-\sigma}(n - \sigma + 1))\}$$
$$+ 2^\sigma \times 2^{n-\sigma}\{(2^{\sigma-1}(2\sigma - 1))(2^{\sigma-1}(2\sigma - 1))\}$$

$$\sum_{\sigma=2}^{n} Sz_v(BB^*(n)/\xi_\sigma)$$
$$= \frac{2^n}{108}\{(31976 - 14856n + 1521n^2 + 288n^3)2^{2n}$$
$$+ (360n^3 + 540n^2 - 6084n - 33696)2^n + 2044\}$$

$$Sz_e(BB^*(n)/\xi_\sigma)$$
$$= 2^{n+2}\{((11 - 12\sigma)(2^{\sigma-1} - 2^{n-1}) + 3(n - \sigma + 1)2^{n-\sigma} - 1)(2^{\sigma-2}(12\sigma - 13)$$
$$+ 3(n - \sigma)2^{n-\sigma}(2^{\sigma+1} - 2) + 2^{n-\sigma}(2^\sigma - 1) + 2^\sigma + 2^{\sigma+1} - 1)\}$$
$$+ 2^{n-\sigma} \times 2^\sigma\{(3(n - \sigma)2^{n-\sigma} + 2^{n-\sigma+1})(3(n - \sigma)2^{n-\sigma} + 2^{n-\sigma+1})\}$$
$$+ 2^\sigma \times 2^{n-\sigma}\{(2^{\sigma-2}(12\sigma - 13) + 2^{\sigma+1})(2^{\sigma-2}(12\sigma - 13) + 2^{\sigma+1})\}$$

$$\sum_{\sigma=2}^{n} Sz_e\, BB^*(n)/\xi_\sigma$$
$$= 2^n\left\{\left(\frac{15383}{6} - \frac{1823n}{2} + \frac{147n^2}{4} + 24n^3\right)2^{2n}\right.$$
$$\left. + \left(30n^3 - \frac{123n^2}{2} - \frac{1445n}{2} - 2784\right)2^n + 4n + \frac{659}{3}\right\}$$

$$Sz_{ev}\, BB^*(n)/\xi_\sigma$$
$$= \frac{1}{2}\Big[2^{n+2}\{(2^{n-\sigma}(n - \sigma + 1) - (2^\sigma - 2^n)(2\sigma - 1))(2^{\sigma-2}(12\sigma - 13)$$
$$+ 3(n - \sigma)2^{n-\sigma}(2^{\sigma+1} - 2) + 2^{n-\sigma}(2^\sigma - 1) + 2^\sigma + 2^{\sigma+1} - 1)$$
$$+ (2^{\sigma-1}(2\sigma - 1) + 2^{n-\sigma}(n - \sigma + 1)(2^{\sigma+1} - 2))\left((11 - 12\sigma)(2^{\sigma-1} - 2^{n-1})\right.$$
$$\left. + 3(n - \sigma + 1)2^{n-\sigma}\right) - 1\}$$
$$+ 2^{n-\sigma} \times 2^\sigma \{2(2^{n-\sigma}(n - \sigma + 1))(3(n - \sigma)2^{n-\sigma} + 2^{n-\sigma+1})\}$$
$$+ 2^{n-\sigma} \times 2^\sigma \{2(2^{\sigma-1}(2\sigma - 1))(2^{\sigma-2}(12\sigma - 13) + 2^{\sigma+1})\}\Big]$$

$$\sum_{\sigma=2}^{n} Sz_{ev}\, BB^*(n)/\xi_\sigma$$
$$= 2^n \left\{ \left( \frac{16235}{18} - \frac{4399n}{12} + \frac{109n^2}{4} + 8n^3 \right) 2^{2n} \right.$$
$$\left. + \left( 10n^3 - \frac{11n^2}{4} - \frac{873n}{4} - 962 \right) 2^n + \frac{572}{9} \right\}$$

$$PI(BB^*(n)/\xi_\sigma) = 2^{n+2}\{((11 - 12\sigma)(2^{\sigma-1} - 2^{n-1}) + 3(n - \sigma + 1)2^{n-\sigma} - 1)$$
$$+ (2^{\sigma-2}(12\sigma - 13) + 3(n - \sigma)2^{n-\sigma}(2^{\sigma+1} - 2) + 2^{n-\sigma}(2^\sigma - 1) + 2^\sigma + 2^{\sigma+1}$$
$$- 1)\} + 2^{n-\sigma} \times 2^\sigma \{(3(n - \sigma)2^{n-\sigma} + 2^{n-\sigma+1}) + (3(n - \sigma)2^{n-\sigma} + 2^{n-\sigma+1})\}$$
$$+ 2^\sigma \times 2^{n-\sigma}\{(2^{\sigma-2}(12\sigma - 13) + 2^{\sigma+1}) + (2^{\sigma-2}(12\sigma - 13) + 2^{\sigma+1})\}$$

$$\sum_{\sigma=2}^{n} PI(BB^*(n)/\xi_\sigma) = 2^n\big((24n^2 - 57n + 82)2^n - 8n - 90\big)$$

$$Mo(BB^*(n)/\xi_\delta)$$
$$= 2^{n+2}\big|(2^{n-\sigma}(n - \sigma + 1) - (2^\sigma - 2^n)(2\sigma - 1))$$
$$- (2^{\sigma-1}(2\sigma - 1) + 2^{n-\sigma}(n - \sigma + 1)(2^{\sigma+1} - 2))\big| + 2^{n-\sigma} \times 2^\sigma \times 0$$
$$+ 2^\sigma \times 2^{n-\sigma} \times 0$$

$$\sum_{\sigma=2}^{n} Mo(BB^*(n)/\xi_\delta) = \sum_{\sigma=2}^{n} 2^{n+2}\big| 2^{n-\sigma}(n - \sigma + 1) + (2\sigma - 1)(2^n - 2^\sigma) - 2^{\sigma-1}(2\sigma - 1)$$
$$- 2^{n-\sigma}(2^{\sigma+1} - 2)(n - \sigma + 1) \big|$$

$$Mo_e(BB^*(n)/\xi_\delta)$$
$$= 2^{n+2}\big|((11 - 12\sigma)(2^{\sigma-1} - 2^{n-1}) + 3(n - \sigma + 1)2^{n-\sigma} - 1)$$
$$- (2^{\sigma-2}(12\sigma - 13) + 3(n - \sigma)2^{n-\sigma}(2^{\sigma+1} - 2) + 2^{n-\sigma}(2^\sigma - 1) + 2^\sigma + 2^{\sigma+1}$$
$$- 1)\big| + 2^{n-\sigma} \times 2^\sigma \times 0 + 2^\sigma \times 2^{n-\sigma} \times 0$$

$$\sum_{\sigma=2}^{n} Mo_e(BB^*(n)/\xi_\delta)$$
$$= \sum_{\sigma=2}^{n} 2^{n+2} \big| (36n - 36\sigma + 16)2^{n-\sigma-2} - (24n - 48\sigma + 26)2^{n-2}$$
$$- (36\sigma - 23)2^{\sigma-2} \big|$$

Hence,

$$Sz_v(BB^*(n)/\xi_\sigma) = Sz_v(BB^*(n)/\xi_1) + \sum_{\sigma=2}^{n} Sz_v(BB^*(n)/\xi_\sigma)$$
$$= \frac{2^n}{54} \{(15988 - 7212n + 882n^2 + 144n^3) \times 2^{2n}$$
$$- (16632 + 3366n - 270n^2 - 180n^3) \times 2^n + 617\}$$

$$Sz_e(BB^*(n)/\xi_\sigma) = Sz_e(BB^*(n)/\xi_1) + \sum_{\sigma=2}^{n} Sz_e(BB^*(n)/\xi_\sigma)$$
$$= 2^n \left\{ \left( \frac{30709}{12} - 922n + 57n^2 + 24n^3 \right) \times 2^{2n} \right.$$
$$\left. - \left( 2756 + \frac{1445}{2}n + \frac{123}{2}n^2 - 30n^3 \right) \times 2^n + 4n + \frac{587}{3} \right\}$$

$$Sz_{ev}\big(BB^*(n)\big) = Sz_{ev}(BB^*(n)/\xi_1) + \sum_{\sigma=2}^{n} Sz_e(BB^*(n)/\xi_\sigma)$$

$$= 2^n \left\{ \left(\frac{16145}{18} - \frac{1087}{3}n + 34n^2 + 8n^3\right) \times 2^{2n} \right.$$
$$\left. - \left(943 + \frac{909}{4}n + \frac{11}{4}n^2 - 10n^3\right) \times 2^n + \frac{410}{9} \right\}$$

$$PI\big(BB^*(n)\big) = PI(BB^*(n)/\xi_1) + \sum_{\sigma=2}^{n} PI(BB^*(n)/\xi_\sigma)$$
$$= 2^n \{(73 - 36n + 24n^2) \times 2^n - 8n - 78\}$$

$$Mo\big(BB^*(n)\big) = Mo(BB^*(n)/\xi_1) + \sum_{\sigma=2}^{n} Mo(BB^*(n)/\xi_\sigma)$$
$$= 2^{n+2} \,|\, n2^{n-1} - 2^n + 3\,|$$
$$+ \sum_{\sigma=2}^{n} 2^{n+2} \,|\, 2^{n-\sigma}(n - \sigma + 1) + (2\sigma - 1)(2^n - 2^\sigma) - 2^{\sigma-1}(2\sigma - 1)$$
$$- 2^{n-\sigma}(2^{\sigma+1} - 2)(n - \sigma + 1)\,|$$

$$Mo_e\big(BB^*(n)\big) = Mo_e(BB^*(n)/\xi_1) + \sum_{\sigma=2}^{n} Mo_e(BB^*(n)/\xi_\sigma)$$
$$= 2^{n+2} \,|\, 2^{n-1} - 3(n-1)2^{n-1} + 2^n - 6\,|$$
$$+ \sum_{\sigma=2}^{n} 2^{n+2} \,|\, (36n - 36\sigma + 16)2^{n-\sigma-2} - (24n - 48\sigma + 26)2^{n-2}$$
$$- (36\sigma - 23)2^{\sigma-2}\,|$$

### 3.3 Numerical Analysis and Graphical Comparison

Using the analytical expressions obtained from the previous section, the structural descriptors for few fixed values of $n$ are computed for the benes network, $BB(n)$ and augmented benes network, $BB^*(n)$. The numerical values of the descriptors for $BB(n)$ are tabulated in Table 2 and Table 3 and that for $BB^*(n)$ are presented in Table 4 and Table 5. Also, the numerical values are plotted graphically and are presented in Figure 13.

Comparing the numerical values of descriptors for $BB(n)$ and $BB^*(n)$ the wiener and szeged index for $BB^*(n)$ is less than that for $BB(n)$ for all n. On the other hand, all the remaining descriptors is high for $BB^*(n)$ compared with that for $BB(n)$ for all n. It is also observed that $Mo_e\big(BB^*(n)\big) = Mo(BB^*(n))$ for $n = 2, 3, 4$.

**Table 2:** Numerical values of distance based descriptors for benes network, $BB(n)$.

| n | Wiener | Szeged | Edge-Szeged | Edge-Vertex-Szeged |
|---|---|---|---|---|
| 2 | 492 | 2912 | 3840 | 3344 |
| 3 | 5944 | 69952 | 129824 | 95392 |
| 4 | 53872 | 1247872 | 2728576 | 1848128 |
| 5 | 412384 | 18711808 | 45143424 | 29116032 |
| 6 | 2823616 | 250456576 | 645950464 | 402967808 |

**Table 3:** Numerical values of distance based descriptors for benes network, $BB(n)$.

| n | Padmakar-Ivan | Mostar | Edge-Mostar |
|---|---|---|---|
| 2 | 736 | 192 | 224 |
| 3 | 7296 | 1408 | 1728 |
| 4 | 54400 | 8448 | 10624 |
| 5 | 350208 | 45568 | 69376 |
| 6 | 2059776 | 250880 | 437760 |

**Table 4:** Numerical values of distance based descriptors for augmented benes network, $BB^*(n)$.

| n | Wiener | Szeged | Edge-Szeged | Edge-Vertex-Szeged |
|---|---|---|---|---|
| 2 | 418 | 1270 | 6148 | 2752 |
| 3 | 5108 | 35628 | 191464 | 82032 |
| 4 | 47016 | 726616 | 4200592 | 1741984 |
| 5 | 365136 | 12043440 | 73545120 | 29729088 |
| 6 | 2531488 | 173578592 | 1105984576 | 438073984 |

**Table 5:** Numerical values of distance based descriptors for augmented benes network, $BB^*(n)$.

| n | Padmakar-Ivan | Mostar | Edge-Mostar |
|---|---|---|---|
| 2 | 1176 | 224 | 464 |
| 3 | 10768 | 1728 | 3808 |
| 4 | 78368 | 10624 | 24000 |
| 5 | 501056 | 58112 | 132992 |
| 6 | 2945152 | 296448 | 683776 |

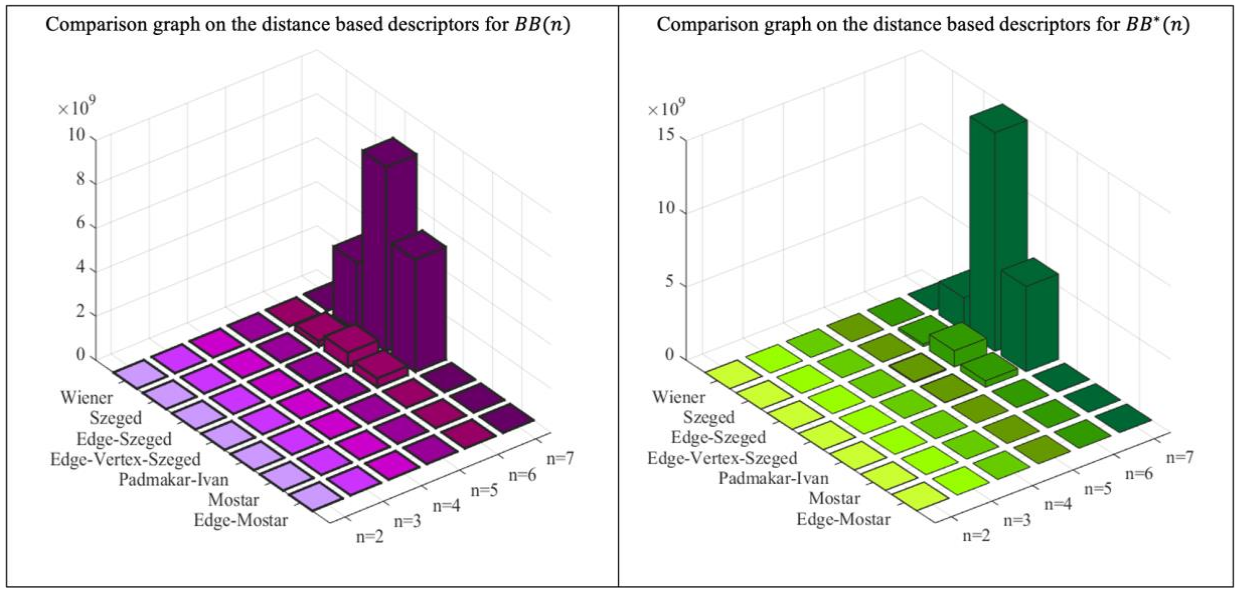

**Figure 13:** Comparison graph on the distance based descriptors for $BB(n)$ and $BB^*(n)$.

## 4 Comparative Analysis

In this section, the topological network parameters to check the efficiency of networks are evaluated for the networks, $BB(n), n \geq 2$ and $BB^*(n), n \geq 2$. The formulas to evaluate various topological network parameters [1] are defined in Table 6.

**Table 6:** Topological network parameters of a graph $\zeta$.

| Network Parameter | Mathematical Expression |
| --- | --- |
| Diameter | $Dia(\zeta) = max\{d_\zeta(\mu, \eta); \mu, \eta \in \mathbb{V}(\zeta)\}$ |
| Average Distance | $W(\zeta) = \sum_{\{\mu,\eta\} \subseteq \mathbb{V}(\zeta)} w_v(\mu) w_v(\eta) d_\zeta(\mu, \eta)$ |
| Message Traffic Density | $MTD(\zeta) = \dfrac{W(\zeta) \times |\mathbb{V}(\zeta)|}{|\mathbb{E}(\zeta)|}$ |
| Network Throughput | $NT(\zeta) = \dfrac{|\mathbb{E}(\zeta)|}{Dia(\zeta)}$ |
| Graph Density | $GD(\zeta) = \dfrac{2|\mathbb{E}(\zeta)|}{|\mathbb{V}(\zeta)|(|\mathbb{V}(\zeta)| - 1)}$ |
| Total Connectivity | $TC(\zeta) = \dfrac{|\mathbb{E}(\zeta)|}{|\mathbb{V}(\zeta)|(|\mathbb{V}(\zeta)| - 1)}$ |

As mentioned, $BB(n), n \geq 2$ has $(2n + 1)2^n$ vertices and $n2^{n+2}$ edges and $BB^*(n), n \geq 2$ has $(2n + 1)2^n$ vertices and $(12n - 1)2^{n-1}$ edges. Using the formulas given in Table 6, the parameters are computed for $BB(n)$, and $BB^*(n)$ and the results are described in Section 4.1 and Section 4.2 respectively.

### 4.1 Results on Benes Network, $BB(n), n \geq 2$.

- Diameter: $Dia(BB(n)) = 2n$
- Average Distance: $W(BB(n)) = \dfrac{2^n}{3}\{(10n^3 - 9n^2 + 50n - 69)2^n + 69\}$
- Message Traffic Density: $MTD(BB(n)) = \dfrac{2^n(2n+1)}{12n}\{(10n^3 - 9n^2 + 50n - 69)2^n + 69\}$
- Network Throughput: $NT(BB(n)) = 2^{n+1}$
- Graph Density: $GD(BB(n)) = \dfrac{8n}{(2n+1)((2n+1)2^n - 1)}$
- Total Connectivity: $TC(BB(n)) = \dfrac{4n}{(2n+1)((2n+1)2^n - 1)}$

### 4.2 Results on Augmented Benes Network, , $BB^*(n), n \geq 2$.

- Diameter: $Dia(BB^*(n)) = 2n$
- Average Distance: $W(BB^*(n)) = \dfrac{2^n}{6}\{(20n^3 - 36n^2 + 148n - 207)2^n + 207\}$
- Message Traffic Density:
  $MTD(BB^*(n)) = \dfrac{2^{n+1}(2n+1)}{6(2n-1)}\{(20n^3 - 36n^2 + 148n - 207)2^n + 207\}$
- Network Throughput: $NT(BB^*(n)) = \dfrac{2^{n-2}(12n-1)}{n}$
- Graph Density: $GD(BB^*(n)) = \dfrac{12n-1}{(2n+1)((2n+1)2^n - 1)}$
- Total Connectivity: $TC(BB^*(n)) = \dfrac{12n-1}{2(2n+1)((2n+1)2^n - 1)}$

## 4.3 Comparative Analysis

Using the expressions obtained in Section 4.1 and Section 4.2, the numerical values of the parameters for a few fixed values of $n$ were determined and graphically plotted. Figure 14 to Figure17 show the graphical representations.

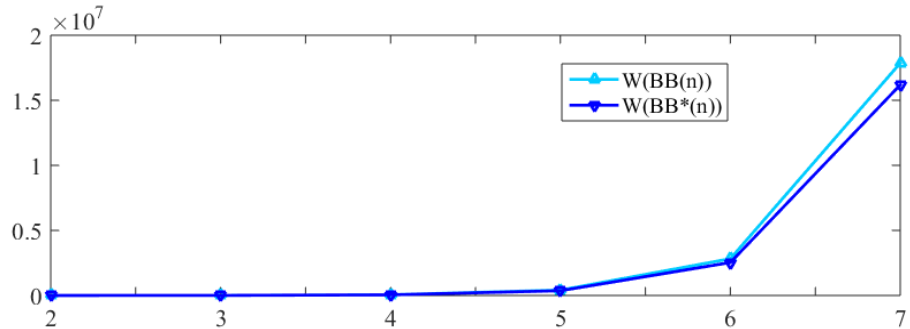

**Figure 13:** Comparison graph based on average distance.

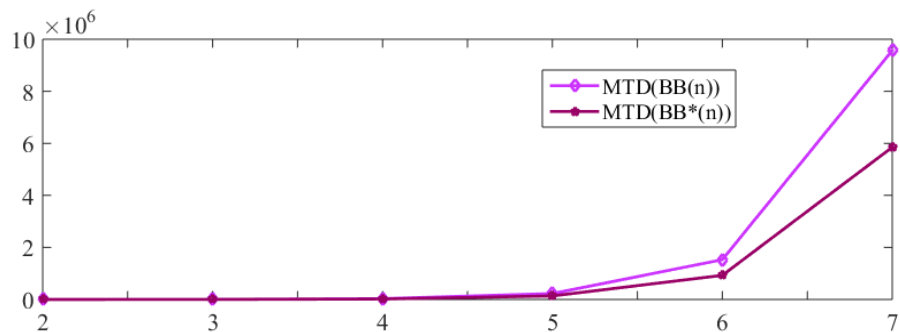

**Figure 15:** Comparison graph based on message traffic density.

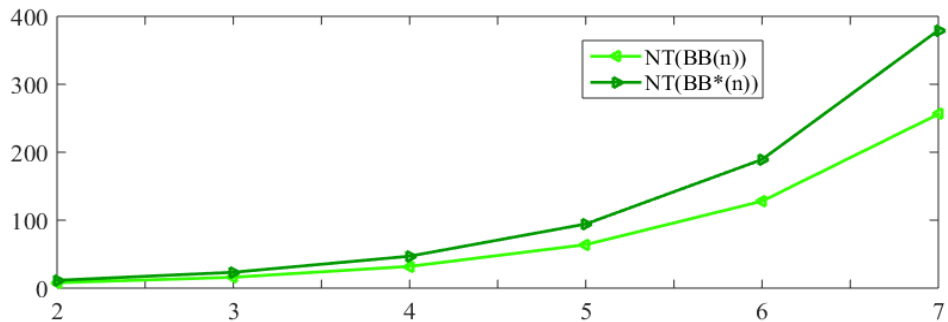

**Figure 16:** Comparison graph based on network throughput.

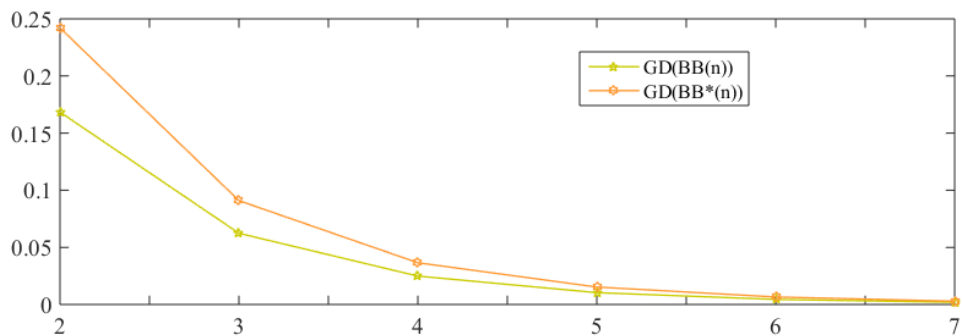

**Figure 17:** Comparison graph based on graph density.

It is to be noted that, the comparison graph based on total connectivity is similar to the comparison graph based on graph density, which is given in Figure 17. From the obtained expressions as well as from the graphical presentations, it can be inferred that $BB^*(n)$, has minimum average distance, low message traffic density, high network throughput, high graph density and high total connectivity, making it more efficient network than the benes network, $BB(n)$.

## 5 Broadcasting Algorithm

Broadcasting is an information dissemination problem in a connected graph, which is a simplest communication problem. In broadcasting process, a node or vertex of a network, called as source node or originating vertex has a message, which is disseminated from the source node to all other nodes, with the condition that each node can only transmit the message to at most one of its neighbors at a time. A spanning tree along which the message is broadcast from the source node to all other nodes is called broadcasting tree of the network. Furthest-Distance-First Protocol is a common strategy used to solve the broadcasting problem, in which, message is broadcasted from the source node to an uninformed adjacent node which leads to longest path in a tree. Broadcasting has various applications in virus spreading, internet messaging and management of distributed systems etc. See [19-27] for details on broadcasting problems on networks and graphs.

*Broadcasting Algorithm for Augmented Benes Network:*

In $BB^*(n)$, each node is labelled as $(j, k)$ where $j$ is a $n$-bit binary number and $0 \leq k \leq 2n$. The eccentricity of a vertex, $\mu \epsilon V(\zeta)$, $e(\eta) = max\{d_\zeta(\mu, \eta)\}$.

1. Input: Let $\mathbb{S}$ be the source node with a message to be broadcasted to all other nodes of the augmented benes network, $BB^*(n)$.
2. Step 1: Construct the broadcasting tree under the conditions
   i. Source node $\mathbb{S}(j, k)$ has either $k = 0$ or $k = 2n$.
   ii. Source node $\mathbb{S}$ has exactly three child nodes.
   iii. The broadcasting tree ends with node whose parent has only two children and the end node $(j, k)$ is such that $k = 0$ or $k = 2n$.
3. Apply Furthest-Distance-First Protocol to disseminate the message along the broadcasting tree.
4. Stop after $e(\mathbb{S}) + 2$ steps, when $n$ is even.
5. Stop after $e(\mathbb{S}) + 3$ steps, when $n$ is odd

*Proof of correctness:*

The source node $\mathbb{S}$ disseminates the message according to Furthest-Distance-First protocol. Hence, $\mathbb{S}$ sends the message by giving priority to an uninformed adjacent node which leads to longest path in the broadcasting tree. Consider $BB^*(2)$, which can possibly have eight source nodes. The broadcasting tree generated for any selection of the source node remains isomorphic to graph $\mathbb{T}$, given in Figure 18. Clearly the message dissemination ends after $e(\mathbb{S})+ 2$ steps, where $e(\mathbb{S}) = 4$. Similarly, the broadcasting tree for $BB^*(3)$ is given in Figure 19, which stops disseminating message after $e(\mathbb{S}) + 3$ steps, where $e(\mathbb{S}) = 6$.

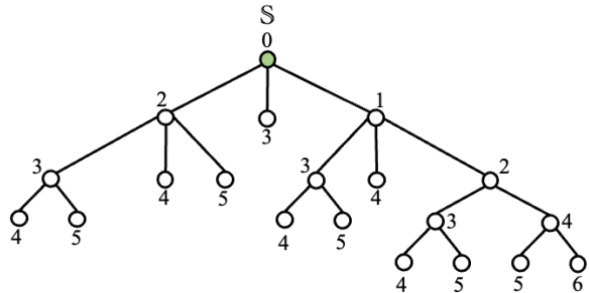

**Figure 18:** Broadcasting tree, $\mathbb{T}$ for $BB^*(2)$ with source node $\mathbb{S}$.

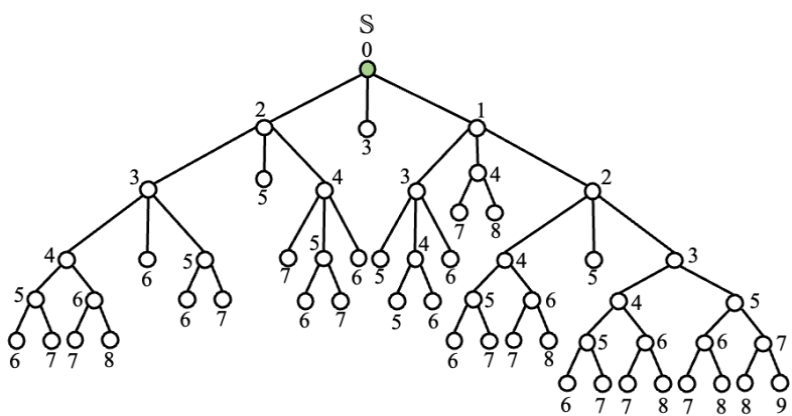

**Figure 19:** Broadcasting tree, $\mathbb{T}$ for $BB^*(2)$ with source node $\mathbb{S}$.

## 6 Conclusions

In this work, a new class of interconnection network is constructed from the benes network, called as the augmented benes network. The distance based descriptors for the n-dimensional benes network as well as for $n$-dimensional augmented benes network were evaluated and the analytical expressions are given in Section 3.1. and Section 3.2 respectively. Further, to recognize the efficiency of the newly derived network compared to the efficiency of benes network, the networks parameters of the networks are computed and a comparative analysis was carried out. The numerical data of the topological descriptors for few dimensions of the benes network and augmented benes network were evaluated from the analytical expressions, with the help of which a graphical comparison was drawn and the same are presented in Section 3.3. This study helps in characterising the topological properties of interconnection networks and helps to design new networks which brings advancements in technology and in field of computer science.

**Declaration:** There is no data availability.

**Conflicts of Interest:** The authors declare that they have no conflicts of interest to report regarding the present study.